\documentclass[pra,twocolumn,amsmath,amssymb,groupedaddress,longbibliography]{revtex4-2}
\usepackage{lineno}
\usepackage{lipsum}
\usepackage{color}
\usepackage{braket}
\usepackage{amsmath}
\usepackage{graphicx}
\usepackage{dcolumn}
\usepackage{appendix}
\usepackage{bm}
\usepackage{orcidlink}
\usepackage{balance} 
\usepackage{float}
\hypersetup{colorlinks=true, linkcolor=red, citecolor=blue, urlcolor=magenta}

\usepackage{lineno}

\begin{document}

\preprint{APS/123-QED}

\title{Correlated dynamics of three-particle bound states induced by emergent impurities in Bose-Hubbard model}
\author{Wenduo Zhao$^{1,2}$}

\author{Boning Huang$^{1,2}$}

\author{Yongguan Ke$^{2}$} \email{keyg@szu.edu.cn}

\author{Chaohong Lee$^{2,3}$}

\affiliation{$^{1}$Laboratory of Quantum Engineering and Quantum Metrology, School of Physics and Astronomy, Sun Yat-Sen University (Zhuhai Campus), Zhuhai 519082, China}

\altaffiliation{Email: chleecn@szu.edu.cn, chleecn@gmail.com}
\affiliation{$^{2}$Institute of Quantum Precision Measurement, State Key Laboratory of Radio Frequency Heterogeneous Integration, College of Physics and Optoelectronic Engineering, Shenzhen University, Shenzhen 518060, China}
\affiliation{$^{3}$Quantum Science Center of Guangdong-Hong Kong-Macao Greater Bay Area (Guangdong), Shenzhen 518045, China}

\date{\today}
\begin{abstract}
Bound states, known as particles tied together and moving as a whole, are profound correlated effects induced by particle-particle interactions.
While dimer-monomer bound states are manifested as a single particle attached to a dimer bound pair, it is still unclear about quantum walks and Bloch oscillations of dimer-monomer bound states.
Here, we revisit three-particle bound states in the Bose-Hubbard model and find that interaction-induced impurities adjacent to bound pair and boundaries cause two kinds of bound states: one is dimer-monomer bound state and the other is bound edge state.
In quantum walks, the spread velocity of dimer-monomer bound state is determined by the maximal group velocity of their energy band, which is much smaller than that in the single-particle case. 
In Bloch oscillations, the period of dimer-monomer bound states is one third of that in the single-particle case.
Emergence of bound edge states also requires that interaction-induced defects are greater than the effective tunneling strength of three-particle bound state. 
Our work provides new insights to basic mechanics and collective dynamics of three-particle bound states. 


\end{abstract}

\maketitle

\section{Introduction}
Interaction between particles causes rich novel correlations and phenomena,
such as bound states~\cite{winkler2006repulsively,Valiente_2008,fukuhara2013microscopic}, i.e. particles tied together and moving as a whole,
has attracted intense attention.
For two particles,
bound states have been theoretically predicted and experimentally observed~\cite{PhysRevA.90.062301,zhong2017floquet,gorlach2017topological,PhysRevA.95.063630,PhysRevB.96.195134,van2019topological,olekhno2020topological,PhysRevA.101.023620,stepanenko2020interaction,ke2020radiative,azcona2021doublons,liu2023correlated,zheng2023two,qin2024occupation,Huang_2024,lei2025topologically}.
These states exist outside the two-particle scattering continuum, separated by large gap comparable to interaction strength.
However, more particles are quite different.
Bound states of three or more particles~\cite{ghirardi1965number,ahmadzadeh1965new,brayshaw1969connection,sitenko1971bound,mattis1984three,macek1986loosely,zouzou1986four,zhen1988loosely,kievsky1994study,brodsky2005bound,gridnev2012zero,meissner2015spectrum,lakaev2017existence,hansen2017applying,meng2018three,liang2018observation,PhysRevA.104.033303,sun2024boundary,lakaev2025existence,liu2025hierarchy}, can exhibit new patterns rather than simple extension of two-particle bound states. 
For example, a dimer–monomer bound state (DMBS) manifested as a single particle attached to adjacent dimer bound pair has been predicted~\cite{PhysRevA.81.011601}.
This is because particle-particle interaction gives rise to effective impurity surrounded by the dimer bound state that traps the single particle.

Bound states can modify quantum dynamics compared to the single-particle case.
In a lattice, a single particle can undergo quantum walks in the absence of external force and Bloch oscillations in the presence of external force.
While quantum walks~\cite{childs2009universal,venegas2012quantum,kadian2021quantum,cai2021multiparticle} can be used to speed up the spread of quantum information~\cite{PhysRevA.48.1687}, Bloch oscillations encode the force information into the oscillation period.
Both of them have been realized in a variety of experimental platforms, such as ions~\cite{PhysRevLett.103.090504,PhysRevLett.104.100503}, cold atoms~\cite{karski2009quantum}, optical waveguide arrays~\cite{perets2008realization}, nuclear magnetic resonance (NMR)~\cite{du2003experimental,ryan2005experimental},  boosting potential applications in quantum information process. 
When extending to two particles, both quantum walks and Bloch oscillations of bound states are featured by strong particle-particle correlations, different to the single-particle counterpart~\cite{preiss2015strongly}.
For example, in quantum walks the spread velocity of bosonic bound state is triple that of fermionic bound state, both slower than that of single particle~\cite{PhysRevA.90.062301}.
In Bloch oscillations two-particle bound state has half period of the single particle~\cite{preiss2015strongly,PhysRevA.99.063614}.
However, the collective dynamical behaviors of DMBSs are essentially unexplored.
\begin{figure}
\centering
\includegraphics[width=1\linewidth]{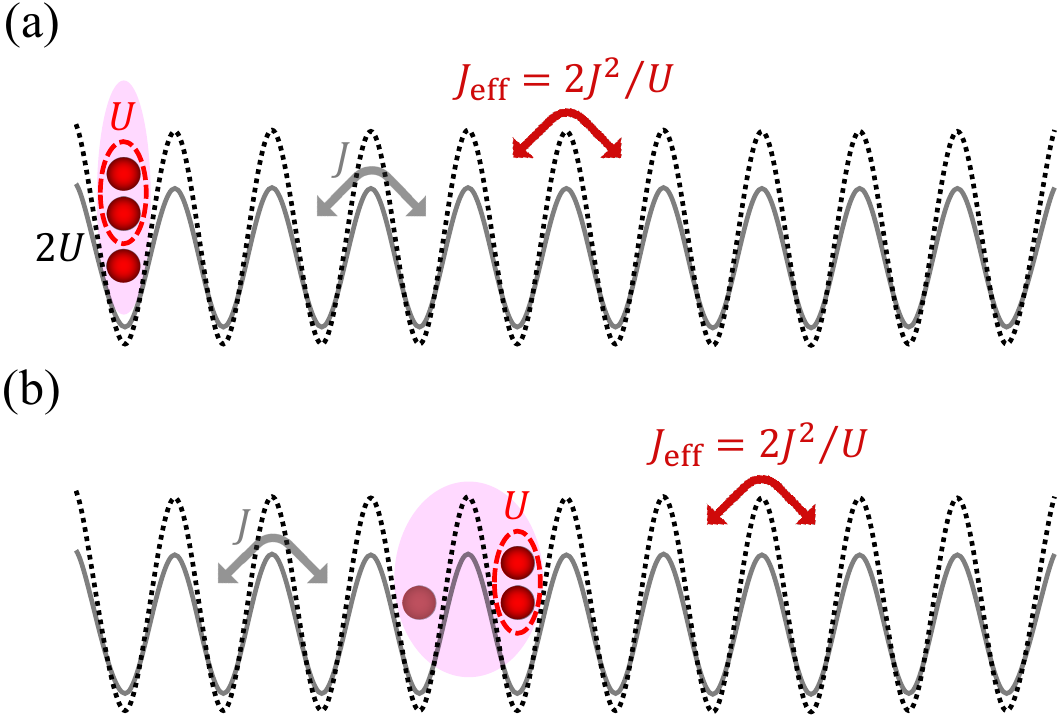}
\caption{Schematics of three particle bound states.
The tunneling strength of single particle is $J$, while the bound dimer pair has effective tunneling strength $2J^2/U$ due to particle-particle interaction $U$.
(a) Three-particle bound edge state due to interaction-induced defect at the boundaries.
(b) dimer-monomer bound state due to interaction-induced defect adjacent to the dimer bound pair.
}
\label{illustration}
\end{figure}
\begin{figure*}
\centering
\includegraphics[width=18cm]{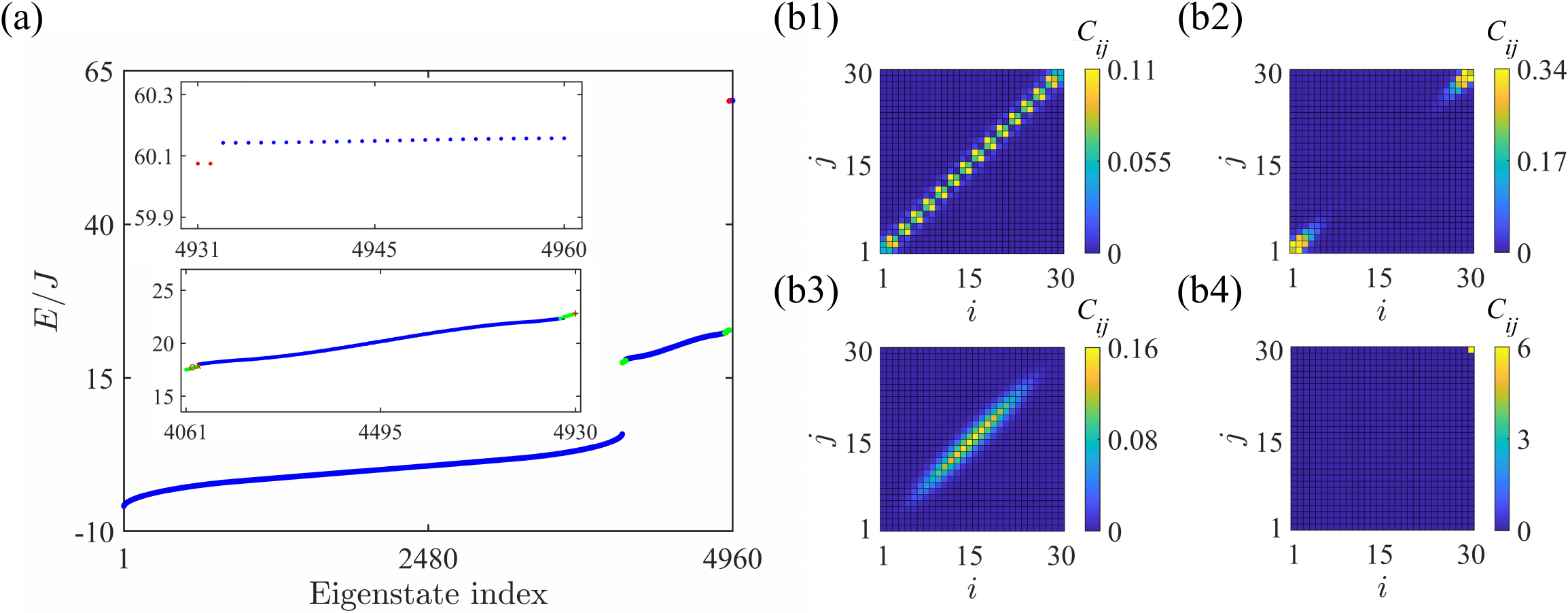}
\caption{Three-particle spectrum and three-particle bound states in the Bose-Hubbard model.
(a) Spectrum of the three-particle Bose-Hubbard model.
The insets are the enlargement of type-(ii)(iii) regions, and the green and red dots marked the dimer-monomer bound states and type-(iii) bound edge states, respectively.
(b1-b3) Second-order correlation functions of three typical dimer-monomer bound states, which are marked by red `o', `x', and `+' in inset of (a).
(b4) Second-order correlation function of type-(iii) bound edge states.
Parameters are set as $J=1$, $U=20$, and $M=30$.
}
\label{threeparticle_pdf}
\end{figure*}

In this work, we study three-particle bound states in Bose-Hubbard model~[as depicted by Fig.~\ref{illustration}], and explore quantum walks and Bloch oscillations of DMBSs. 
Under the strong interaction, three-particle states can be classified into three types of states: (i) scattering states behaving as three independent particles at distinct sites, (ii) dimer-monomer states behaving as combination of a bound pair and an independent particle, and (iii) three-particle bound states behaving as all particles tied at the same site.
DMBSs known as a dimer tied by a single particle come from type-(ii) states, while bound edge states known as all three correlated particles localized at the boundaries come from type-(iii) states.
To understand mechanics of the two special states, we apply approach of Chrieffer–Wolff
transformation to obtain the effective Hamiltonian for DMBSs, which is more accurate than the one obtained by adiabatically
eliminating method~\cite{PhysRevA.81.011601}. 
With the effective Hamiltonian, we find that there are two kinds of interaction-induced defects: one is at the boundaries and the other is adjacent to the dimer bound pair in the bulk. 
The emergent boundary defects trap all three particles at the boundaries, explaining the origin of bound edge state; see Fig.~\ref{illustration}(a). 
Bound edge state is different from topological edge states originating from nontrivial bulk topology and the traditional defect edge states originating from real defects at the boundaries.
With a simplified two-component Hubbard model, we find that the emergence of bound edge state requires that the interaction-induced defects are greater than the effective tunneling of three-particle bound state.  
Two-particle bound edge state does not exist because the interaction-induced defects are equal to the effective tunneling of two-particle bound state.
The bulk defect around the dimer pair traps the other single particle, explaining the origin of DMBSs; see Fig.~\ref{illustration}(b). 
Importantly, we also find DMBS localized at the boundaries due to the joint effect of the two kinds of emergent defects.
We further study key features of DMBSs in quantum dynamics such as quantum walks and Bloch oscillations.
In quantum walks, we find that the spread velocity is determined by the maximal group velocity of the bands of DMBS.
In Bloch oscillations, we find the period is one third of that of single-particle.
Our work provides new insights to the three-particle bound states and collective correlated dynamics.

\section{Three-particle bound states} \label{SecII}
We consider the Bose-Hubbard model described by the Hamiltonian
\begin{eqnarray}
    \hat{H}=-\sum_{j=1}^{M-1}J(\hat{b}_{j}^{\dag}\hat{b}_{j+1}+{\rm H.c.})+\frac{U}{2}\sum_{j=1}^{M}\hat{n}_{j}(\hat{n}_{j}-1),
    \label{eq1}
\end{eqnarray}
where $\hat{b}_{j}^{\dag}$ ($\hat{b}_{j}$) is the bosonic creation (annihilation) operator at $j$th site, and $\hat{n}_j=\hat{b}_{j}^{\dag}\hat{b}_{j}$ is the particle number operator. 
The system size is $M$ and open boundary condition is adopted here.
$J$ is the nearest-neighbor tunneling strength, 
and $U$ is the strength of onsite interaction between particles.
When the interaction strength is strong ($U\gg J$), particles at the same site will form bound states, which tunnel as a whole. 
Diagonalizing Hamiltonian~\eqref{eq1} in the three-particle subspace, we obtain the energy spectrum [Fig.~\ref{threeparticle_pdf}(a)], which is clearly divided into three regions corresponding to (i) three-particle scattering state, (ii) two-particle bound state and one independent particle, and (iii) three-particle bound state from low to high energies, respectively.
Here, the parameters are set as $J=1$, $U=20$, and $M=30$.

We concentrate on the type-(ii) and type-(iii) states, where three-particle bound states can be formed.
For the type-(ii) case, there are states that the dimer bound pair is always adjacent to the other particle, termed as dimer-monomer bound states (DMBSs) and marked by green color in Fig.~\ref{threeparticle_pdf}(a).
Figs.~\ref{threeparticle_pdf}(b1,b2,b3) show second-order correlation functions $C_{ij}=\bra{\psi}\hat{b}_{i}^{\dag}\hat{b}_{j}^{\dag}\hat{b}_{j}\hat{b}_{i}\ket{\psi}$ of three typical DMBSs.
The dominant distributions on the same site and adjacent site indicate featured correlation of the DMBSs.
For the type-(iii) case, all the three particles are tied at the same site.
Fig.~\ref{threeparticle_pdf}(b4) shows a type-(iii) bound edge state which behaves as three particles tied together at the boundary; marked by red dots in Fig.~\ref{threeparticle_pdf}(a).
One can distinguish DMBSs and bound edge states in experimental platform of ultracold atoms in optical lattice~\cite{greiner2002quantum,preiss2015strongly}.
While DMBSs have extended density profile in the bulk and dominant first and second diagonal lines in correlation function, bound edge states have localized density profile at the boundary and dominant diagonal corner in correlation function. 
Both density distribution and density-density correlation can be accessible via site-resolved imaging in quantum gas microscope~\cite{preiss2015strongly,gross2017quantum}. 
Compared states in Figs.~\ref{threeparticle_pdf}(b2) and (b4), we find that both of them are edge states with different localized lengths.
Later, it will be clear that these edge states can be explained by interaction-induced defects at the boundaries.

\begin{figure}[h]
\centering
\includegraphics[width=1\linewidth]{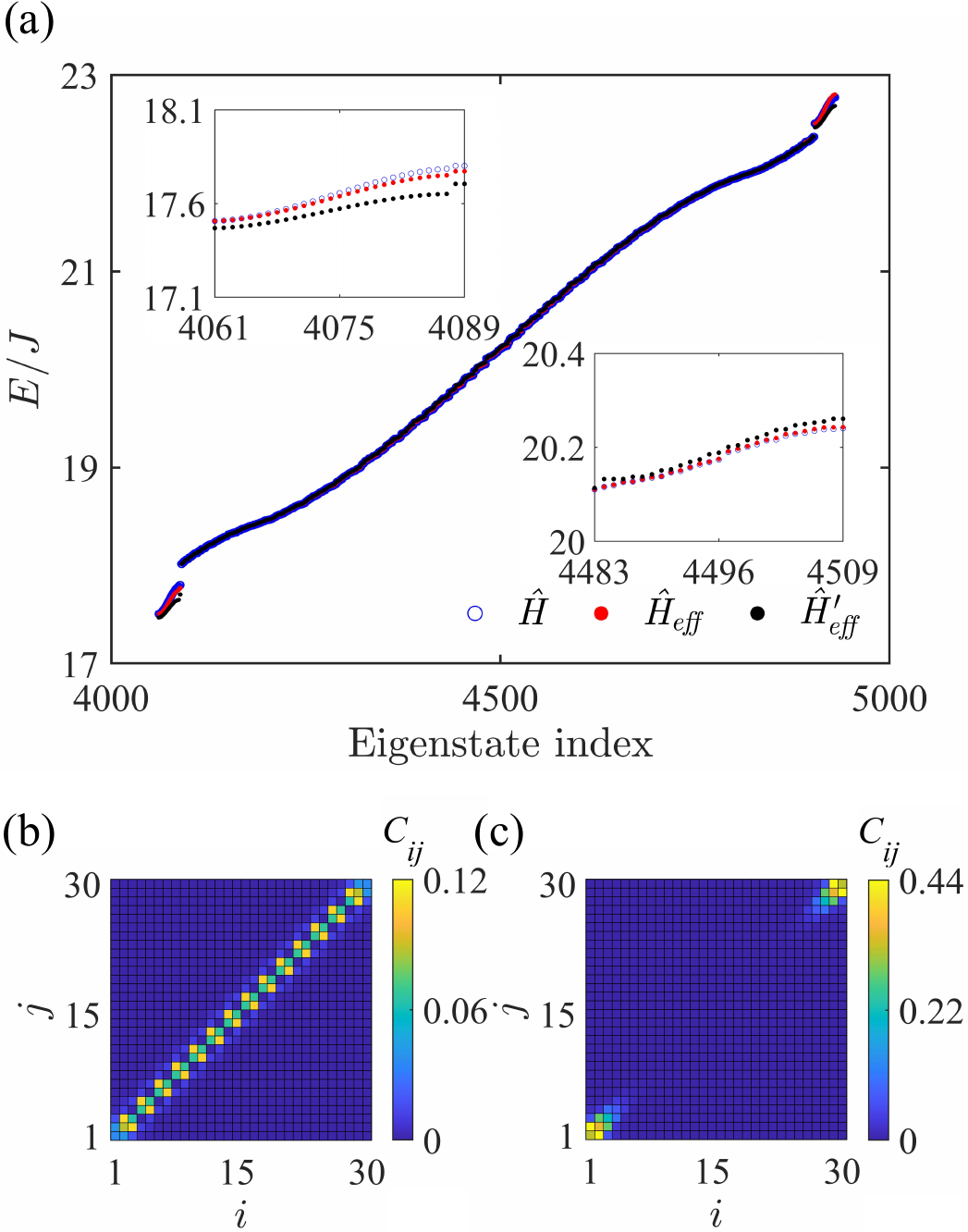}
\caption{The validity of the effective model $\hat{H}_{\rm eff}$.
(a) Eigenvalues correspond to type-(ii) states given by the original model (blue circles), the effective model $H_{\rm{eff}}$ (red dots) and the effective model $H_{\rm{eff}}'$ in Ref.~\cite{PhysRevA.81.011601} (black dots).
Insets are the enlargement for parts of the eigenvalues.
(b) and (c) are second-order correlation functions of two typical three-particle bound states obtained from the effective Hamiltonian $\hat{H}_{\rm eff}$.
Parameters are set as $J=1$, $U=20$, and $M=30$.
}
\label{Eeff_3p_pdf}
\end{figure}

The DMBSs in Figs.~\ref{threeparticle_pdf}(b1-b3) originate from interaction-induced emergent defects. 
To understand the formation of DMBSs, we can obtain an effective Hamiltonian for type-(ii) states via the perturbation theory~\cite{MTakahashi_1977,bravyi2011schrieffer}.
When $U\gg J$, the interaction term 
\begin{eqnarray}
    \hat{H}_{U}=\frac{U}{2}\sum_{j=1}^{M}\hat{n}_{j}(\hat{n}_{j}-1)
    \label{eq2}
\end{eqnarray}
is dominant, and the hopping term 
\begin{eqnarray}
    \hat{H}_J=-\sum_{j=1}^{M-1}J(\hat{b}_{j}^{\dag}\hat{b}_{j+1}+{\rm H.c.})
    \label{eq3}
\end{eqnarray}
can be treated as a perturbation. 
In the three-particle subspace, eigenstates $|2\rangle_i|1\rangle_j$, $(i\neq j)$ of $\hat{H}_U$ with eigenvalues $E_{(iij)}=U$ form a subspace, 
while eigenstates $|1\rangle_i|1\rangle_j|1\rangle_k$, ($i\neq j\neq k$) with $E_{(ijk)}=0$ and $|3\rangle_j$ with $E_{(jjj)}=3U$ form a complemental subspace.
Here, $|2\rangle_i|1\rangle_j=(1/\sqrt{2})(\hat{b}_{i}^{\dag })^2\hat{b}_{j}^{\dag}\ket{0}$, $|1\rangle_i|1\rangle_j|1\rangle_k=\hat{b}_{i}^{\dag }\hat{b}_{j}^{\dag }\hat{b}_{k}^{\dag}\ket{0}$, and $|3\rangle_j=(1/\sqrt{6})(\hat{b}_{j}^{\dag })^3\ket{0}$ are Fock states with $\ket{0}$ being the vacuum state.
We respectively define the projection operators on the two subspaces as
\begin{eqnarray}\label{eq4}
    &&\hat{P}=\sum_{i\neq j}|2\rangle_i|1\rangle_j\langle1|_j\langle2|_i,\\ 
&&\hat{S}=\sum_{i\neq j \neq k}\frac{|1\rangle_i|1\rangle_j|1\rangle_k\langle1|_i\langle1|_j\langle1|_k}{E_{(iij)}-E_{(ijk)}}+\sum_{j}\frac{|3\rangle_j\langle3|_j}{E_{(iij)}-E_{(jjj)}}. \nonumber 
\end{eqnarray}
Applying the degenerate perturbation theory up to second order, the effective Hamiltonian of type-(ii) states is given by
\begin{eqnarray}
    \hat{H}_{\rm eff}=\hat{P}\hat{H}\hat{P}+\hat{P}\hat{H}_{J}\hat{S}\hat{H}_{J}\hat{P}.
    \label{eq5}
\end{eqnarray}
Substituting Eqs.~(\ref{eq2},~\ref{eq3},~\ref{eq4}) into Eq.~\eqref{eq5} and simplifying it,
one can obtain the effective Hamiltonian for the subspace spanned by $\{|2\rangle_i|1\rangle_j\}$,
\begin{widetext}
\begin{eqnarray}  
    \hat{H}_{\rm eff}&&=\sum_{j}\big(\frac{2J^{2}}{U}\hat{c}_{j}^{\dag}\hat{c}_{j+1}-J\hat{b}_{j}^{\dag}\hat{b}_{j+1}+{\rm H.c.}\big)+\sum_{i\ne j}(U+\epsilon_{i,j})\hat{c}_{i}^{\dag}\hat{c}_{i}\hat{b}_{j}^{\dag}\hat{b}_{j}+\sum_{\left|i-j\right|=1,2}S_{|i-j|}\hat{c}_{i}^{\dag}\hat{c}_{j}\hat{b}_{j}^{\dag}\hat{b}_{i} \nonumber  \\ 
    &&+\sum_{j}\bigg[(\frac{2J^{2}}{U}\hat{c}_{j+1}^{\dag}\hat{c}_{j}+\frac{2J^{2}}{U}\hat{c}_{j}^{\dag}\hat{c}_{j-1}+\frac{J^{2}}{2U}\hat{c}_{j}^{\dag}\hat{c}_{j})\hat{b}_{j-1}^{\dag}\hat{b}_{j+1}+{\rm H.c.}\bigg] \label{eq6} 
\end{eqnarray}
Here, $\hat{c}_j^\dagger\equiv (1/\sqrt{2} )\hat{b}_{j}^{\dag}\hat{b}_{j}^{\dag}\ket{0}$ and $\hat{b}_j^\dagger$ are the creation operators of dimer bound pair and a monomer, respectively.
The first term indicates the nearest-neighbor tunnelings of dimer bound pair with strength $2J^{2}/U$ and monomer with strength $-J$, respectively.
The second term indicates the site-dependent energy $\epsilon_{i,j}$ which follows
\begin{equation}
    \epsilon_{i,j}=
    \begin{cases}
    4J^{2}/U,~\text{for}~ \left|i-j\right|>1,~i\in \rm bulk; \\
    J^{2}/(2U),~\text{for}~ \left|i-j\right|=1,~i\in \rm bulk;\\
    2J^{2}/U,~\text{for}~\left|i-j\right|>1,~i\in \rm edge; \\
    -3J^{2}/(2U), ~\text{for}~\left|i-j\right|=1,~i\in \rm edge, 
    \end{cases}
\end{equation}
where $i\in {\rm edge}$ means $i=1$ or $i=M$, and $i\in {\rm bulk}$ means $i\neq1$ and $i\neq M$.
The energy difference between bulk and boundary site gives rise to effective defects at the boundaries, which can explain the DMBS localized at the boundaries; see Fig.~\ref{threeparticle_pdf}(b2).
The energy difference between two kinds of DMBSs $|2\rangle_{i}|1\rangle_{j}$ with ($|i-j|=1$) and ($|i-j|>1$) is comparable to $J^2/U$, which can explain finite energy gap between DMBSs and other type (ii) states.
The third term indicates the swap of dimer and monomer with strengths $S_1=-2J$ for dimer-monomer distance equal to $1$  and $S_2=2J^2/U$ for dimer-monomer distance equal to $2$.
The swap process with strength $S_1$ couples DMBSs $|2\rangle_j|1\rangle_{j+1}$ and $|1\rangle_j|2\rangle_{j+1}$.
That is why the energy difference between two bands of DMBSs [marked by green dots in Fig.~\ref{threeparticle_pdf}] is comparable to $J$.
The fourth term indicates nearest-neighbor tunneling of monomer assisted by states of the dimer.
When the dimer-monomer distance is equal to $2$, the dimer is shifted by one site in the opposite direction of monomer. 
When the dimer-monomer distance is equal to $1$, 
the dimer is shifted by one site in the opposite direction of monomer or stays in the same site.

Note that Ref.~\cite{PhysRevA.81.011601} has  obtained a similar effective Hamiltonian
\begin{equation}
    \begin{aligned}   
    \hat{H}_{\rm eff}' &=\sum_{j}\bigg(\frac{2J^{2}}{U}\hat{c}_{j}^{\dag}\hat{c}_{j+1}-J\hat{b}_{j}^{\dag}\hat{b}_{j+1}+{\rm H.c.}\bigg)+\sum_{i\ne j}(U+\epsilon'_{i,j})\hat{c}_{i}^{\dag}\hat{c}_{i}\hat{b}_{j}^{\dag}\hat{b}_{j}+\sum_{\left|i-j\right|=1}S_{1}\hat{c}_{i}^{\dag}\hat{c}_{j}\hat{b}_{j}^{\dag}\hat{b}_{i},
    \label{eqA1}
    \end{aligned}
\end{equation}
\end{widetext}
where the on-site energy $\epsilon_{i,j}'$ is given by $\epsilon_{i,j}'=4J^{2}/U$ for $\left|i-j\right|>1$ and 
$\epsilon_{i,j}'=J^{2}/(2U)$ for $\left|i-j\right|=1$, and the swap of dimer and monomer with strengths $S_1=-2J$ for dimer-monomer distance equal to $1$.
Compared Eq.~\eqref{eq6} with Eq.~\eqref{eqA1}, 
our effective Hamiltonian maintains more second-order perturbation terms including defects at the boundaries,  next-nearest-neighbor swap of dimer and monomer,  and the dimer-assisted next-nearest-neighbor tunneling of monomer.  
%
%
Our effective Hamiltonian is more accurate than that in the Ref.~\cite{PhysRevA.81.011601}.
To show this, we calculate the energy spectrum of the effective Hamiltonians $\hat{H}_{\rm eff}$, $\hat{H}_{\rm eff}'$ and the spectrum of original model; see  Fig.~\ref{Eeff_3p_pdf}.
Parameters are chosen as $J=1$, $U=20$, and $M=30$.
It is clear that the spectrum of our effective Hamiltonian is overall closer to the spectrum of type-(ii) states of the original Hamiltonian.
Without loss of generality, we show the second-order correlation of the DMBSs obtained from our effective Hamiltonian $\hat{H}_{\rm eff}$, corresponding to Figs.~\ref{threeparticle_pdf} (b1,b2).
They exhibit almost the same distribution.
These numerical results demonstrate our  effective Hamiltonian provides a good description of the type-(ii) states when interaction is strong. 
However, as interaction strength decreases to $U=0.5$ and further to $U=0.05$, the DMBS with lower energy will first vanish and finally all DMBSs are destroyed; see Appendix~\ref{Appendix:B}.

\section{Dynamics of dimer–monomer bound states}
The correlated bound states will lead to different dynamical properties to single-particle scenario.
As the type-(iii) bound states is only the simple extension of onsite two-particle bound states, whose dynamical properties are well-known, we concentrate on the type-(ii) bound states, especially the DMBSs.
We have already known the static properties of DMBSs.
To further identify DMBSs, we rely on quantum walks without external force and Bloch oscillations with external force, which are closely related to the band structure. 
Since the energy bands of DMBSs are isolated from the other type-(ii) bands, we can obtain a simple tight-binding model to understand the dynamics related to the bands of DMBSs.

To achieve this, we first adopt period boundary condition to calculate the center-of-mass Bloch band~\cite{PhysRevA.95.063630}.
According to the co-translational symmetry,  the center-of-mass momentum $\kappa$ of particles become a good quantum number, and the multi-particle Bloch states $|\psi_n(\kappa)\rangle$ with eigenvalue $E_{n,\kappa}$ can be obtained.
Consequently, three-particle Wannier states can be constructed as a Fourier transformation of three-particle Bloch states in the  band of DMBSs
\begin{eqnarray}
    |W_{R}\rangle=\frac{1}{\sqrt{M}}\sum_{\kappa}e^{-i\kappa R}\ket{\psi_{n}(\kappa)},
    \label{WannierState}
\end{eqnarray}
where the Wannier state centered at $R$th site.
As there is a degree of freedom in the phase of Bloch states, the Wannier state can be maximally localized~\cite{PhysRevB.56.12847}.
Here, we obtain the maximally localized Wannier states corresponding to the type-(ii) bound states by using the projection position operator conveniently~\cite{huang2024interaction}.
At last, we can obtain a tight-binding model based on the three-particle Wannier states
\begin{equation} \label{Hw}
    H_{w}=\sum_{R}J_{w} |W_{R}\rangle \langle W_{R+1}|+{\rm H.c.}+E_{w} |W_{R}\rangle \langle W_{R}|,
\end{equation}
where $J_{w}=1/M\sum_\kappa e^{-i\kappa} E_{n,\kappa}$ is the effective tunneling strength and $E_{w}=1/M\sum_\kappa  E_{n,\kappa}$ is the effective onsite energy of Wannier state, which can be neglected.
In the following, we can understand the dynamical  properties of DMBSs in quantum walks and Bloch oscillations based on the Hamiltonian $H_w$.

\begin{figure}[h]
\centering
\includegraphics[width=1\linewidth]{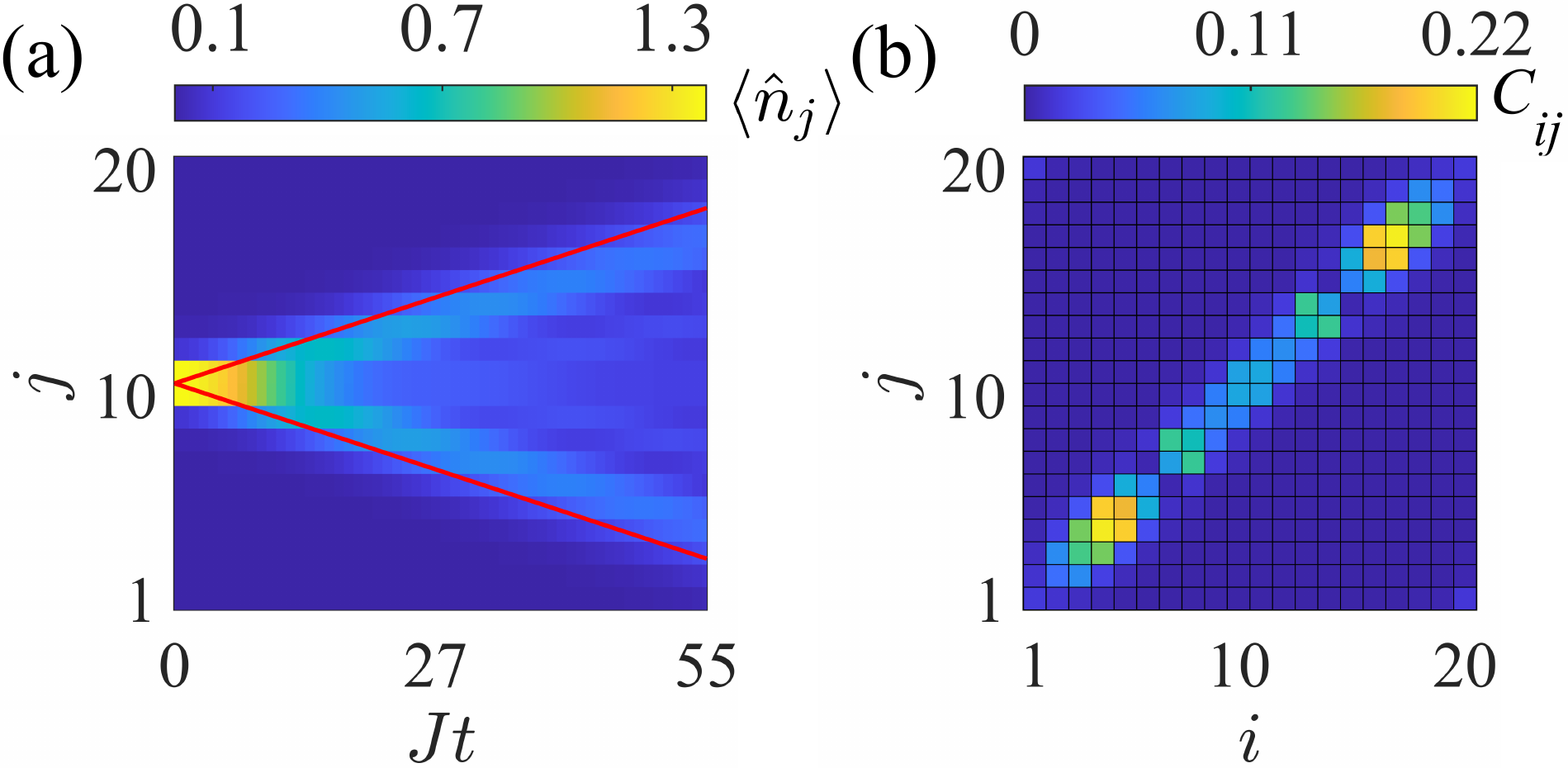}
\caption{Quantum walks of dimer-monomer bound state. (a) Density distribution as a function of time. (b) Second-order correlation of the evolved state at the end time. The red lines in (a) denote propagation trajectory predicted by the group velocity. Parameters are set as $J=1$, $U=20$, $M=20$.}
\label{quantum_walk_pdf}
\end{figure}

\subsection{Quantum walks}
Setting a maximally localized Wannier state corresponding to a band of DMBSs as the initial state, we consider quantum walks governed by $|\psi(t)\rangle=e^{-i\hat{H}t}|\psi(0)\rangle$, as shown in Fig.~\ref{quantum_walk_pdf}(a), where the density distribution $\langle \hat{n}_j \rangle=\bra{\psi(t)}\hat{n}_j\ket{\psi(t)}$ evolves as the time. 
Similarly to quantum walks of a single particle, quantum walks of DMBSs also behave like a ballistic trajectory.
Because the initial state is in the superposition of Bloch states with different group velocities,
the dispersion of wave-packet come results from different group velocities.   
The outmost contours can be determined by the maximal group velocity, which is given by $ vt=\frac{dE}{d\kappa}t$.
According to the tight-binding Hamiltonian~\eqref{Hw}, the energy band is approximated as $E\approx 2J_w \cos(\kappa)$, and hence $vt\approx \pm 2 J_w t$. 
In Fig.~\ref{quantum_walk_pdf}(a),  the red solid line denote the distance predicted by the $\pm 0.14t$,  which maps well to the outmost contours.
Because of small effective hopping strength for the type-(ii) bound states, the group velocities are also small, leading to the slower propagation compared to the single-particle case.
We also show the second-order correlation functions of the evolved state at the end time; see Fig.~\ref{quantum_walk_pdf}(b).
The particles always keep the correlation pattern featured by a dimer bound pair tied to adjacent to the other particle during the dynamics.
\begin{figure}
\centering
\includegraphics[width=1\linewidth]{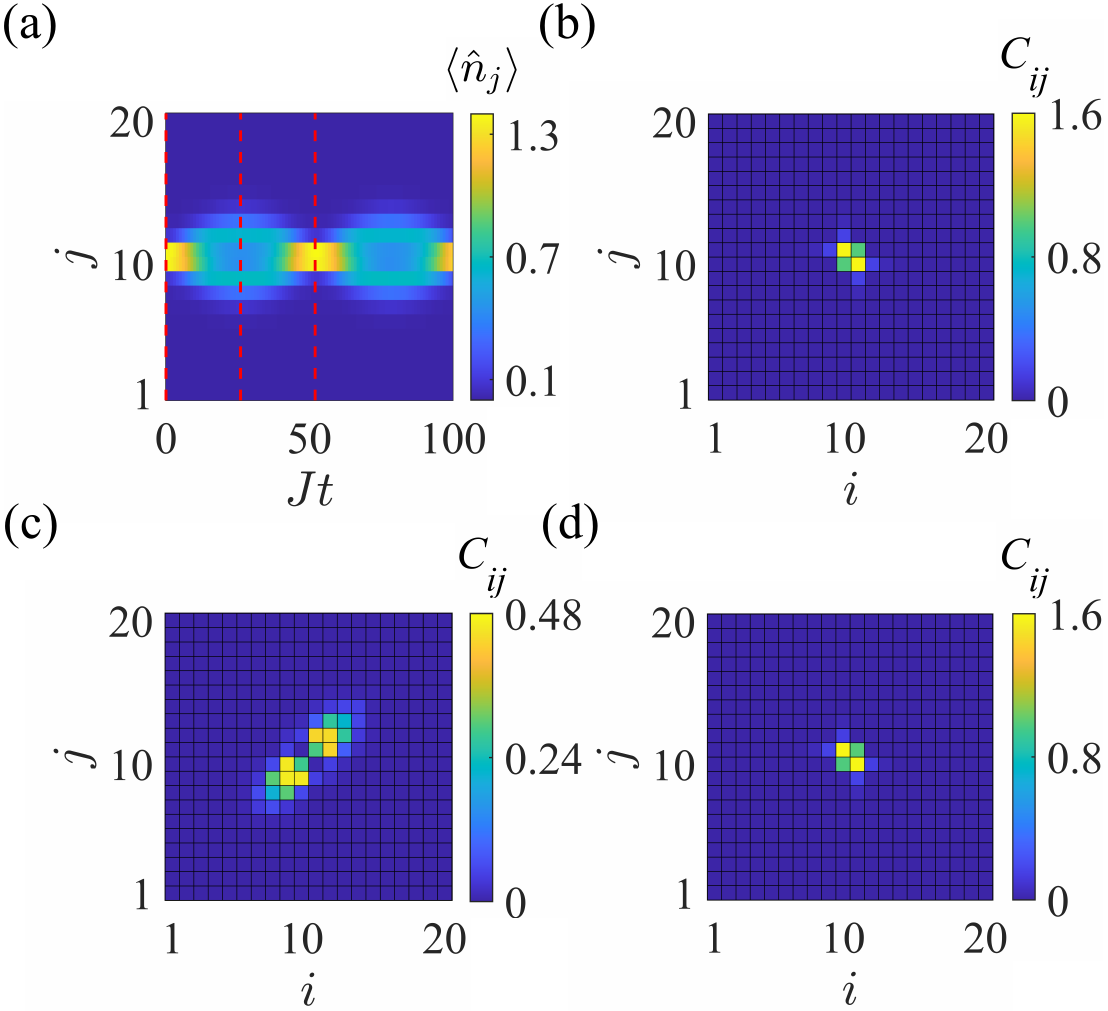}
\caption{Bloch oscillations of dimer-monomer bound state.
(a) Density distribution as a function of time.
(b)(c)(d) Second-order correlation functions of the evolved state at three moments marked by red lines in (a), respectively.
Parameters are set as $J=1$, $U=20$, $F=0.04$ and $M=20$.
}
\label{bloch_oscilation_pdf}
\end{figure}

\subsection{Bloch oscillations}
Another diagnostic of dynamical properties is Bloch oscillations.
Under a static force a single particle will undergo Bloch oscillations~\cite{Hartmann_2004}.
Here, we add an external tilt potential term into the Hamiltonian, $\hat{H}_F=\hat{H}+F\sum_{j=1}^Mj\hat{n}_j$,
and calculate dynamics $|\psi(t)\rangle=e^{-i\hat{H}_Ft}|\psi(0)\rangle$ of the same initial state in the case of quantum walks; see Fig.~\ref{bloch_oscilation_pdf}(a). 
The parameters are set as $J=1$, $U=20$, $F=0.04$ and $M=20$.
Figs.~\ref{bloch_oscilation_pdf}(b-d) show the second-order correlation of the evolved state at three moments $Jt=0$, $Jt=26$ and $Jt=52$, respectively.
During the dynamical process, the bound pair is always tied to the adjacent single particle, and the DMBS undergoes a breathing mode as a whole.
The correlation function at the period $2\pi/{3F}$ returns to the correlation function at the initial time; see Figs.~\ref{bloch_oscilation_pdf}(b) and (d).
We can understand the period of Bloch oscillations and the interval of wave-packet in space by using an approximated tight-binding model,
\begin{equation}
    H_{F,\rm{eff}}\approx \sum_{R}J_{w} |W_{R}\rangle \langle W_{R+1}|+{\rm H.c.}+3FR |W_{R}\rangle \langle W_{R}|,
\end{equation}
where we have neglected the overall onsite energy $E_w$.
This equation is similar to the one describes Bloch oscillations of a single particle~\cite{Hartmann_2004}.
We can immediately obtain the period of oscillations as $2\pi/3F$ and the maximal width of wave-packet in space as $8J_{w}/3F$. 
The small spatial spread interval can be explained by small effective hopping strength $J_w$ and the triple external force.

In dynamics of both quantum walks and Bloch oscillations, the characteristic time scale is $Jt\sim 100$.
In a typical experimental setup of ultracold atoms (e.g. Ref.~\cite{greiner2002quantum,preiss2015strongly}), the tunneling strength is tunable up to the order of $kHz$.
$Jt=100$ could correspond to experimental time as  hundreds of milliseconds, well within the coherence time of current atomic experiments.
Hence, both quantum walks and Bloch oscillations of DMBBs can be readily observed in experiments of ultracold atoms.

\section{Intuitive picture of bound edge states}
As last, we try to understand the origin of 
edge mode of type-(iii) bound states in Fig.~\ref{threeparticle_pdf}(b4).
Motivated by the effective Hamiltonian for type-(ii) bound states, we can treat the single particle and dimer bound pair as particles $A$ and $B$.
While $A$ and $B$ are not possibly occupied in the same site in the previous effective Hamiltonian, we relieve this restriction and introduce onsite interaction between particles $A$ and $B$.
To simplify the picture, we consider a two-component Hubbard model,
\begin{eqnarray}
    \hat{H}_{2}=-\sum_{v,j}J_{v}(\hat{b}_{j}^{(v)\dag}\hat{b}_{j+1}^{(v)}+\text{H.c.})+V\sum_{i=1}^{M} \hat{n}_j^{(A)}\hat{n}_j^{(B)},
    \label{eq7}
\end{eqnarray}
where $\hat{b}_{j}^{(v)\dag}$ ($\hat{b}_{j}^{(v)}$) is the creation (annihilation) operators of type-$v$ ($v\in A,B$) particles at the $j$th site, respectively.
$n_j^{(v)}=\hat{b}_{j}^{(v)\dag}\hat{b}_{j}^{(v)}$ is the density operator of type-$v$ particle.
$J_{v}$ are the nearest-neighbor tunneling strength of type-$v$ particle.
%
%
$V$ is the onsite interaction strength.
Similarly, the effective Hamiltonian of the two-particle bound state can be obtained.
The hopping term 
\begin{eqnarray}
    \hat{H}_1&=&-\sum_{v,j}J_{v}(\hat{b}_{j}^{(v)\dag}\hat{b}_{j+1}^{(v)}+\rm{H.c.})
    \label{EM1D2PP}
\end{eqnarray}
is treated as a perturbation to the interaction term 
\begin{eqnarray}
    \hat{H}_0=V\sum_{i=1}^{M} \hat{n}_j^{(A)}\hat{n}_j^{(B)}.
    \label{EM1D2PM}
\end{eqnarray}
There are degenerate eigenstates $|A_j\rangle|B_j\rangle$ of $\hat{H}_{0}$ with eigenvalues $E_2=V$ forming the subspace of two-particle bound states,
and degenerate eigenstates $|A_i\rangle|B_j\rangle$ ($i\neq j$) with eigenvalues $E_1=0$ forming the complemental subspace.
The projection operators on two subspaces are respectively expressed as
\begin{equation}
\begin{aligned}
&\hat{P}'=\sum_{j=1}^{M}|A_j\rangle|B_j\rangle \langle A_j|\langle B_j|,\\
&\hat{S}'=\sum_{i\neq j}\bigg(\frac{1}{E_{2}-E_{1}}\bigg)|A_i\rangle|B_j\rangle \langle A_i|\langle B_j|.
\end{aligned}
\end{equation}
Using the degenerate perturbation theory up to second order, we can get the effective Hamiltonian of the two-particle bound states as
\begin{eqnarray}
    \hat{H}_{2,\rm eff}=\sum_{j}\big(\frac{2J_{A}J_{B}}{V}\hat{d}_{j}^{\dag}\hat{d}_{j+1}+\text{H.c.}\big)+\varepsilon_{j} \hat d_j^\dagger \hat d_j,
    \label{EM1D2PR}
\end{eqnarray}
where we introduce creation operator of a trimer, $\hat{d}_{j}^{\dag}\ket{0}= \hat{b}_{j}^{(A)\dag}\hat{b}_{j}^{(B)\dag}\ket{0}$, and the onsite energy is given by  $\varepsilon_{j=\{1,M\}}=V+{(J_{A}^{2}+J_{B}^{2})}/{V}$ at the boundaries and   $\varepsilon_{j\ne \{1,M\}}=V+2(J_{A}^{2}+{J_B^{2})}/{V}$ in the bulk.
For strong interaction ($V\gg J_{A}, J_B$), the effective Hamiltonian $\hat{H}_{2,\text{eff}}$ correctly describes bound edge states obtained by $\hat{H}_2$; see Appendix~\ref{Appendix:A}.

\begin{figure}[!htp]
\centering
\vspace{10pt} 
\includegraphics[width=1\linewidth]{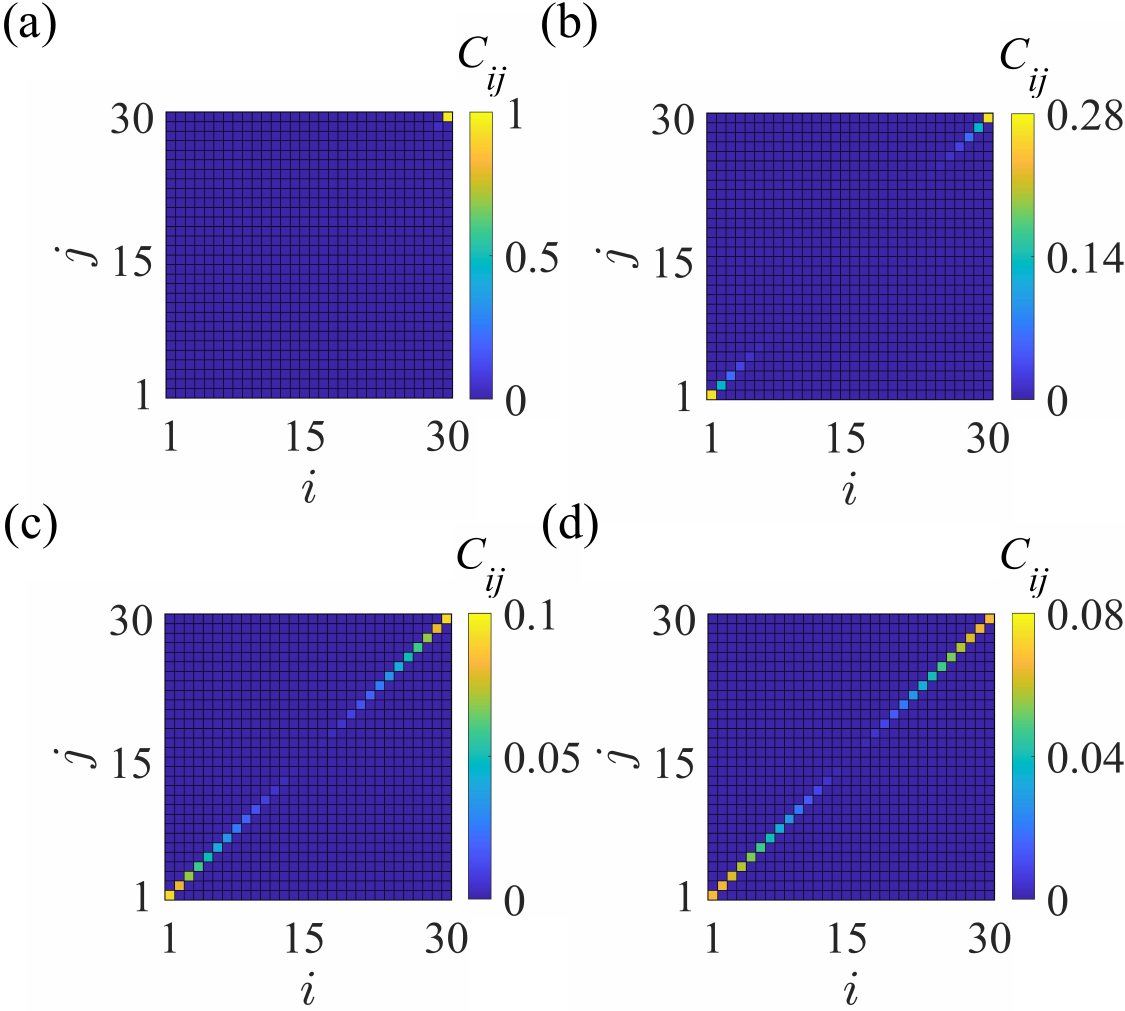}
\caption{Second-order correlation function of two-particle bound edge states with (a) $J_{B}/J_{A}=0.1$, (b) $J_{B}/J_{A}=0.4$, (c) $J_{B}/J_{A}=0.7$ and (d) $J_{B}/J_{A}=1$.
Other parameters are set as $J_A=1$, $V=40$, $M=30$.
}
\label{twoparticles_t1_pdf}
\end{figure}

Clearly, there is an energy difference $(J_A^{2}+J_B^{2})/V$ between the bulk and edge sites, which can be viewed as interaction-induced defects.
The boundary defects lead to the bound edge states.
Fig.~\ref{twoparticles_t1_pdf} shows the correlation function of bound edge states for different $J_B/J_A$, which is defined as $C_{ij}=\bra{\psi}\hat{b}_{j}^{(A)\dag}\hat{b}_{j}^{(A)}\hat{b}_{i}^{(B)\dag}\hat{b}_{i}^{(B)}\ket{\psi}$.
We can find that when $J_B/J_A$ is much smaller or larger than $1$, the bound edge states become stronger localization.
When $J_B/J_A=1$, the state becomes almost extended.
To explain this, we define the ratio between the boundary defect and effective tunneling as
\begin{equation}
    r=\frac{(J_A^{2}+J_B^{2})/V}{{2J_{A}J_{B}}/V}=\frac{1}{2}(\frac{J_A}{J_B}+\frac{J_B}{J_A}).
\end{equation}
The ratio $r$ measures the strength of relative defect.
Larger ratio $r$ will lead to stronger localization.
When $J_A=J_B$, the ratio reaches its minimal value as $1$ and the localization vanishes, because the defect is not strong enough to trap the bound state at the boundaries.
%

%
%


%

Although we treat this problem under strong interaction strength, the analytical results give new insight for bound edge states even when interaction is comparable to the hopping strength.
As interaction decreases, the bound edge states persist localization at the boundaries with energy resided in the continuum of scattering states, which can be termed as the three-particle bound state in continuum~\cite{sun2024boundary}.
Note that Ref.~\cite{sun2024boundary} has already shown that such interaction-induced bound state in continuum exists when there are at least $3$ particles.
The emergence of general bound edge states also requires the same condition, which can be understood in the physical picture mentioned above. 
For two particles, the hopping strength is the same $J_A=J_B=J$. Although there is still interaction-induced defect, the effective tunneling of bound state is the same as the interaction-induced defect, which does not support bound edge states.
However, for three particles, because the hopping strengths of the bound pair $J_B=2J^2/U$ and the other particle $J_A=J$ are always different,
the interaction-induced defect is greater than the effective tunneling of three-particle bound state, which makes the three-particle bound state trap in the boundaries.

\section{Summary and discussions} 

We have revealed two kinds of interaction-induced emergent defects in the three-particle Bose-Hubbard model, which can induce
DMBSs and bound edge states.
In the bulk, quantum walks and Bloch oscillations can be used to to diagnose the features of DMBSs.
The DMBS collectively moves with reduced tunneling in quantum walks, and undergoes one third period of Bloch oscillations compared to the single-particle case.
At the boundaries, the bound edge states can be explained by boundary defects originated from the difference of effective tunnelings of dimer and monomer.

It is interesting to study the interplay between particle-particle correlation and topology in the near future.
Previous works have revealed topological two-particle bound state~\cite{PhysRevB.96.195134,PhysRevA.101.023620,gorlach2017topological,olekhno2020topological,stepanenko2020interaction,ke2020radiative,azcona2021doublons,zheng2023two}, Thouless pumping of two-particle bound states~\cite{PhysRevA.95.063630,PhysRevA.101.023620,tao2025emulating}, topological pumping of three-particle bound state in continuum~\cite{huang2024interaction}.
There may be no counterpart to the single-particle topological states that can be well theorized by the single-particle topological band. 
However, DMBS as a new kind of correlation pattern, may bring new insight into the correlated topological states.

Quantum statistics plays a crucial role in quantum walks of interacting particles~\cite{PhysRevA.90.062301,cai2021multiparticle,kwan2024realization}. 
Spread velocities of bound states are affected by fermionic and bosonic statistics~\cite{PhysRevA.90.062301,cai2021multiparticle}. 
Unlike symmetric spread of fermionic and bosonic particles,
interaction between anyons can change the spread behavior to asymmetry~\cite{kwan2024realization}. 
Furthermore, anyonic statistics can induce  synthetic gauge fields for topological flat bands~\cite{zhang2023anyonic,zhou2025anyonic}.
It is interesting to study how the anyonic statistics affects static and nonequilibrium behaviors of DMBSs.
We believe there are unexplored richness to understand the crossover from few-particle to many-particle topological states.

%

\appendix
\section{Numerical results of the two-component Hubbard model} \label{Appendix:A}
\begin{figure}
\centering
\includegraphics[width=1\linewidth]{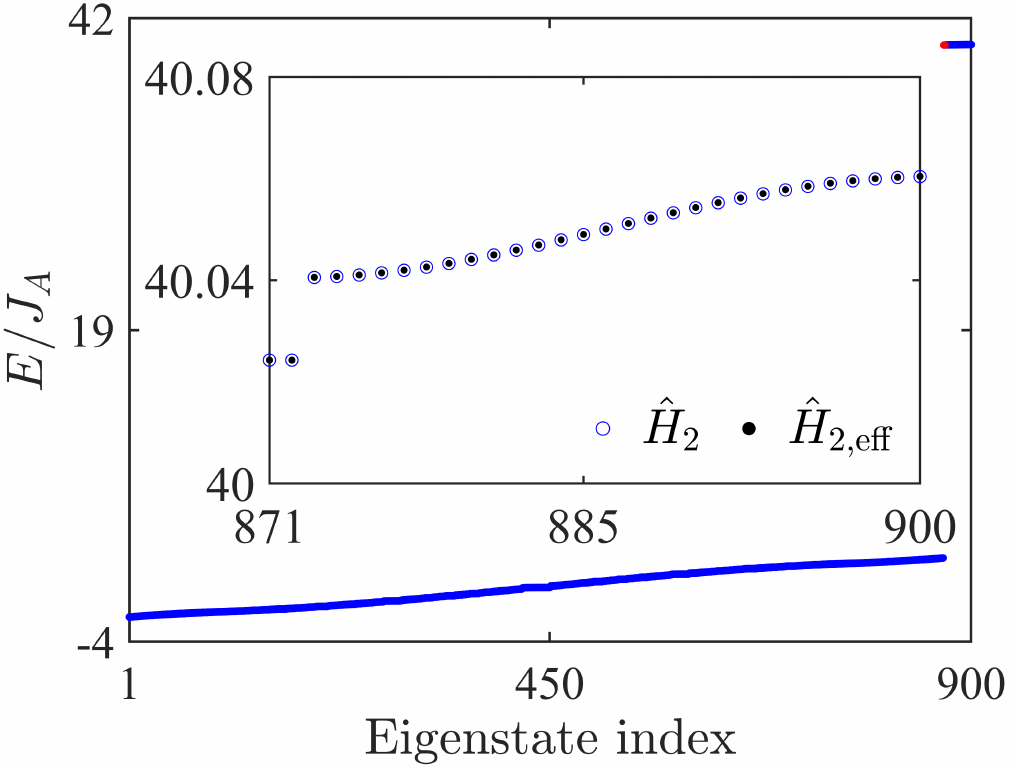}
\caption{Energy spectrum of the two-component Hubbard model. Inset shows the comparison between the original Hamiltonian (blue circles) and the effective Hamiltonian of two-particle bound states (black dots).
Parameters are set as $J_B/J_A=0.1$, $J_A=1$, $V=40$, $M=30$. 
}
\label{twoparticles_pdf}
\end{figure}
We verify the effectiveness of $H_{2,\rm eff}$ to describe bound edge states in the two-component Hubbard model $\hat{H}_{2}$.
Fig.~\ref{twoparticles_pdf} shows the energy spectrum of $\hat{H}_{2}$, which is divided into two regions, and they correspond to the two-particle scattering states and bound states from low to high energies, respectively.
Parameters are set as $J_A=1$, $J_B=0.1$, $V=40$, $M=30$. 
The inset in Fig.~\ref{twoparticles_pdf} shows the comparison between $\hat{H}_{2}$ and the effective Hamiltonian of two-particle bound states $\hat{H}_{2,\rm eff}$.
They are consistent with each other, indicating that $\hat{H}_{2,\rm eff}$ well captures the physics of bound states.

\section{The role of interaction in the formation of DMBSs} \label{Appendix:B}
\begin{figure}
\centering
\includegraphics[width=1\linewidth]{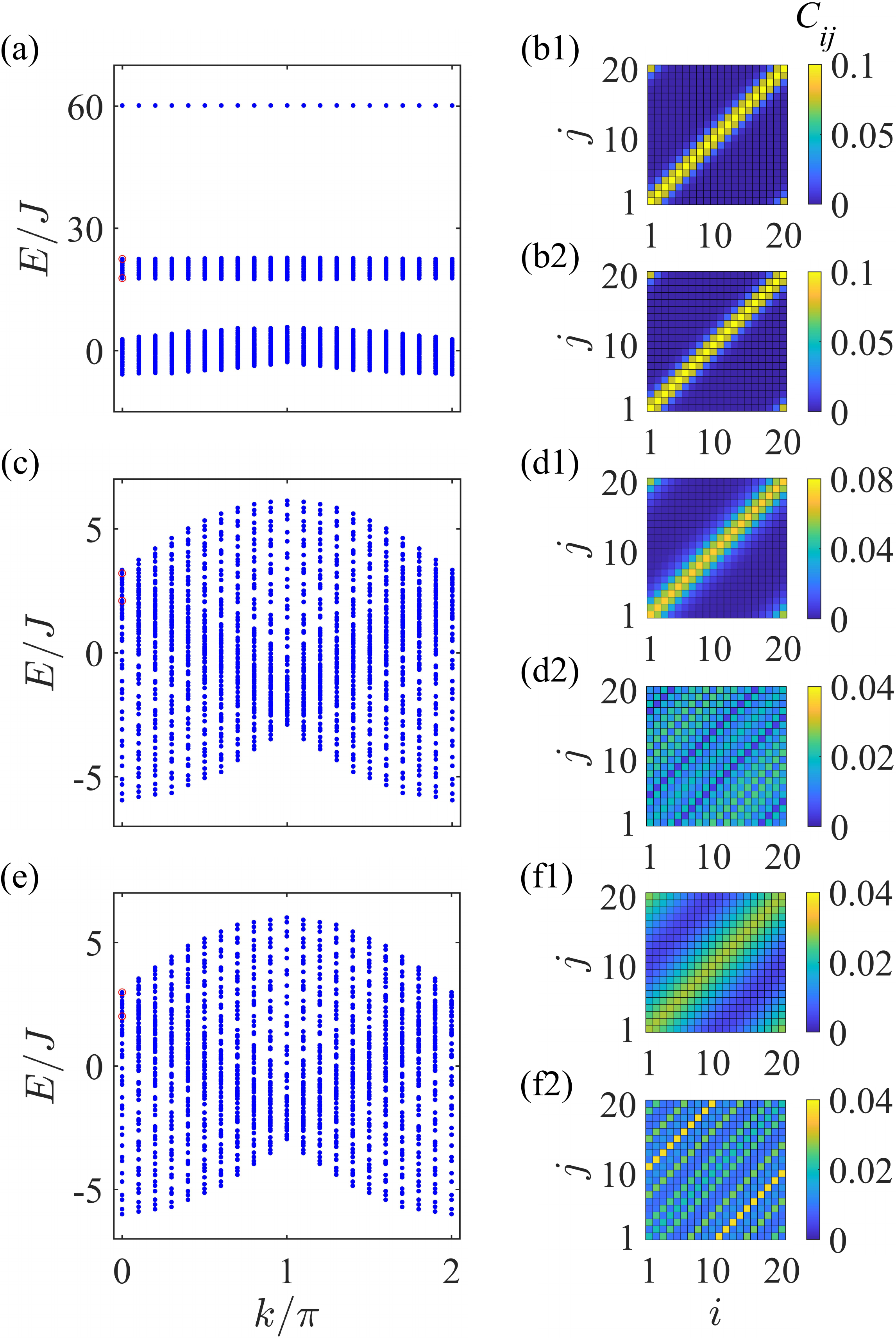}
\caption{Three-particle center-of-mass bands and the $76$th and $58$th eigenstates under different interaction strengths in the Bose-Hubbard model.
(a, c, e): center-of-mass bands under $U=20,~0.5,~0.05$, respectively. 
(b1, b2), (d1, d2), (f1, f2): the $76$th and $58$th eigenstates under $U=20,~0.5,~0.05$, respectively.
Other parameters set as $J=1$, and $M=20$.
}
\label{energy_k_u_pdf}
\end{figure}
In the main text, we show the DMBSs in the case of strong interaction.
DMBSs originate from interaction-induced defects adjacent to the bound pair.
Apparently, if the interaction decreases $0$, the DMBSs will disappear.
It is interesting to understand how the DMBSs disappear as the interaction decreases.
Adopting our previous method~\cite{PhysRevA.95.063630}, under periodic boundary condition we calculate center-of-mass (c.m.) bands with different interaction strengths.
Parameters are set as $J=1$ and $M=20$.
For strong interaction $U=20$, the three types of states are separated into three clusters of c.m. bands; see Figs.~\ref{energy_k_u_pdf}(a).
The $76$th and $58$th three-particle Bloch states with zero c.m. momentum are DMBSs that are in the top and bottom of the second clusters of c.m. bands (marked by red circles in c.m. bands); see Figs.~\ref{energy_k_u_pdf}(b1-b2). 
We trace the $76$th and $58$th Bloch states with zero c.m. momentum when the interaction is decreased to $U=0.5$ and $U=0.05$.
For $U=0.5$, the energy bands are mixed with each other; see Figs.~\ref{energy_k_u_pdf}(c).
However, the $76$th three-particle Bloch state with zero c.m. momentum is still DMBS, while the $58$th eigenstate with zero c.m. momentum becomes scattering state; see Figs.~\ref{energy_k_u_pdf}(d1-d2).
For $U=0.05$, the weak interaction almost plays no role, and the energy bands are also mixed with each other; see Figs.~\ref{energy_k_u_pdf}(e).
All eigenstates including the above $76$th and $58$th become scattering states; see Figs.~\ref{energy_k_u_pdf}(f1-f2).
From the above three cases, we can conclude that interaction plays a crucial role in the formation of these DMBSs.

\nocite{*}
%


\end{document}